\documentclass[pre,twocolumn]{revtex4}

\usepackage{amssymb}
\usepackage{graphicx}

\begin{document}

\title{\textbf{Dynamics towards the Feigenbaum attractor}}

\author{A. Robledo}
\affiliation{Instituto de F\'{\i}sica, Universidad Nacional Aut\'onoma de M\'exico, 
Apartado postal 20-364, M\'exico 01000 D.F., M\'exico}
\author{L. G. Moyano}
\affiliation{Departamento de Matem\'aticas and Grupo Interdisciplinar de Sistemas Complejos,
Universidad Carlos {\rm III} de Madrid, 28911 Legan\'es, Madrid, Spain}

\begin{abstract}
We expose at a previously unknown level of detail the features of the
dynamics of trajectories that either evolve towards the Feigenbaum attractor
or are captured by its matching repellor. Amongst these features are the
following: i) The set of preimages of the attractor and of the repellor are
embedded (dense) into each other. ii) The preimage layout is obtained as the
limiting form of the rank structure of the fractal boundaries between
attractor and repellor positions for the family of supercycle attractors.
iii) The joint set of preimages for each case form an infinite number of
families of well-defined phase-space gaps in the attractor or in the
repellor. iv) The gaps in each of these families can be ordered with
decreasing width in accord to power laws and are seen to appear sequentially
in the dynamics generated by uniform distributions of initial conditions. v)
The power law with log-periodic modulation associated to the rate of
approach of trajectories towards the attractor (and to the repellor) is
explained in terms of the progression of gap formation. vi) The relationship
between the law of rate of convergence to the attractor and the
inexhaustible hierarchy feature of the preimage structure is elucidated.


\end{abstract}
\maketitle

\section{Introduction}

In the last few years there has been a reappraisal and further exploration
of the special dynamical properties displayed by critical attractors in
simple low-dimensional maps \cite{robledo1}. This increased attention has
been induced in part by the perception that the exploration of possible
limits of validity of the canonical statistical mechanics can benefit from
the study of much simpler dynamical systems that are known to exhibit
statistical-mechanical analogies \cite{beck1}. For this purpose in mind an
ideal model system is a one-dimensional map at the transition between
chaotic and regular behavior, represented by well known critical attractors,
such as the Feigenbaum attractor \cite{beck1} \cite{schuster1}. So far,
analytical studies have concentrated on the dynamics \textit{inside} the
attractor, characterized by fluctuating, memory-preserving, nonmixing,
phase-space trajectories. These studies have revealed that these
trajectories obey remarkably rich scaling properties. The results are exact
and clarify \cite{robledo2} the relationship between the original
modification \cite{politi1} \cite{mori1} of the thermodynamic approach to
chaotic attractors \cite{thermo1} \cite{thermo2} \cite{thermo3} for this
type of incipiently chaotic attractor, and some aspects of the $q$%
-statistical formalism \cite{robledo3} that have been seen to manifest for
the same critical attractors. Formerly, dynamics inside the attractor was
not known at this level of detail \cite{robledo3}. The complementary part of
the dynamics, that of advance\textit{\ on the way }to the attractor, has, to
our knowledge, not been analyzed, nor understood, with similar degree of
thoroughness. The process of convergence of trajectories into the Feigenbaum
attractor poses several interesting questions that we attempt to answer here
and elsewhere based on the comprehensive new knowledge presented below.
Prominent amongst these questions is the nature of the connection between
the two sets of dynamical properties, within and outside the attractor.

Trajectories inside the attractor visit positions forming oscillating
deterministic patterns of ever increasing amplitude. However, when the
trajectories are observed only at specified times, positions align according
to power laws, or $q$-exponential functions that share the same $q$-index
value \cite{robledo2} \cite{robledo3}. Further, all such sequences of
positions can be shifted and seen to collapse into a single one by a
rescaling operation similar to that observed for correlations in glassy
dynamics, a property known as `aging' \cite{robledo3} \cite{robledo4}. The
structure found in the dynamics is also seen to consist of a family of
Mori's $q$-phase transitions \cite{mori1}, via which the connection is made
between the modified thermodynamic approach and the $q$-statistical property
of the sensitivity to initial conditions \cite{robledo2} \cite{robledo3}. On
the other hand, a foretaste of the nature of the dynamics outside the
critical attractor can be appreciated by considering the dynamics towards
the so-called supercycles, the family of periodic attractors with Lyapunov
exponents that diverge towards minus infinity. This infinite family of
attractors has as accumulation point the transition to chaos, which for the
period-doubling route is the Feigenbaum attractor. As described in Ref. \cite%
{moyano1}, the basins of attraction for the different positions of the
cycles develop fractal boundaries of increasing complexity as the
period-doubling structure advances towards the transition to chaos. The
fractal boundaries, formed by the preimages of the repellor, display
hierarchical structures organized according to exponential clusterings that
manifest in the dynamics as sensitivity to the final state and transient
chaos. The hierarchical arrangement expands as the period of the supercycle
increases \cite{moyano1}.

Here we present details on the general procedure followed by trajectories to
reach the Feigenbaum attractor, and its complementary repellor. We consider
an ensemble of uniformly distributed initial conditions $x_{0}$ spanning the
entire phase space interval. This is a highly structured process encoded in
sequences of positions shared by as many trajectories with different $x_{0}$%
. There is always a natural dynamical ordering in the $x_{0}$ as any
trajectory of length $t$ contains consecutive positions of other
trajectories of lengths $t-1$, $t-2$, etc. with initial conditions $%
x_{0}^{\prime }$, $x_{0}^{\prime \prime }$, etc. that are images under
repeated map iterations of $x_{0}$. The initial conditions form two sets,
dense in each other, of preimages of each the attractor and the repellor.
There is an infinite-level structure within these sets that, as we shall
see, is reflected by the infinite number of families of phase-space gaps
that complement the multifractal layout of both attractor and repellor.
These families of gaps appear sequentially in the dynamics, beginning with
the largest and followed by other sets consisting of continually increasing
elements with decreasing widths. The number of gaps in each set of
comparable widths increases as $2^{k}$, $k=0,1,\ldots$ and their widths can be
ordered according to power laws of the form $\alpha ^{-k}$, where $\alpha $
is Feigenbaum's universal constant $\alpha \simeq 2.5091$. We call $k$ the
order of the gap set. Furthermore, by considering a fine partition of phase
space, we determine the overall rate of approach of trajectories towards the
attractor (and to the repellor). This rate is measured by the fraction of
bins $W(t)$ still occupied by trajectories at time $t$ \cite{lyra1}. The
power law with log-periodic modulation displayed by $W(t)$ \cite{lyra1} is
explained in terms of the progression of gap formation, and its self-similar
features are seen to originate in the unlimited hierarchy feature of the
preimage structure.

Before proceeding to expand our description in the following sections we
recall \cite{schuster1} the general definition of the interval lengths or
diameters $d_{N,m}$ that measure the bifurcation forks that form the
period-doubling cascade sequence in the logistic map $f_{\mu }(x)=1-\mu x^{2}
$, $-1\leq x\leq 1$, $0\leq \mu \leq 2$. These quantities are measured when
considering the superstable periodic orbits of lengths $2^{N}$, i.e. the $%
2^{N}$-cycles that contain the point $x=0$ at $\overline{\mu }_{N}<\mu
_{\infty }$, where $\mu _{\infty }=1.401155189\ldots$ is the value of the
control parameter $\mu $ at the period-doubling accumulation point \cite%
{beck1}. The positions of the limit $2^{\infty }$-cycle constitute the
Feigenbaum attractor. The $d_{N,m}$ in these orbits are defined (here) as
the (positive) distances of the elements $x_{n}$, $m=0,1,2,\ldots,2^{N}-1$, to
their nearest neighbors $f_{\overline{\mu }_{N}}^{(2^{N-1})}(x_{m})$, i.e. 
\begin{equation}
d_{N,m}\equiv \left\vert f_{\overline{\mu }_{N}}^{(m+2^{N-1})}(0)-f_{%
\overline{\mu }_{N}}^{(m)}(0)\right\vert .
\end{equation}%
For large $N$, $d_{N,0}/d_{N+1,0}\simeq \alpha $. Notice that infinitely
many other sequences of superstable attractors appear at the period-doubling
cascades within the windows of periodic attractors for values of $\mu >$ $%
\mu _{\infty }$. At each of these period-doubling accumulation points occur
replicas of the Feigenbaum attractor. We present explicit results for the
logistic map, that has a quadratic maximum, but the results are easily
extended to unimodal maps with general nonlinearity $z>1$.

Central to our discussion is the following broad property: Time evolution at 
$\mu _{\infty }$ from $t=0$ up to $t\rightarrow \infty $ traces the
period-doubling cascade progression from $\mu =0$ up to $\mu _{\infty }$.
Not only is there a close resemblance between the two developments but also
quantitative agreement. For instance, the trajectory inside the Feigenbaum
attractor with initial condition $x_{0}=0$, the $2^{\infty }$-supercycle
orbit, takes positions $x_{t}$ such that the distances between appropriate
pairs of them reproduce the diameters $d_{N,m}$ defined from the supercycle
orbits with $\overline{\mu }_{N}<\mu _{\infty }$. See Fig. 1, where the
absolute value of positions and logarithmic scales are used to illustrate
the equivalence. This property has been key to obtain rigorous results for
the sensitivity to initial conditions for the Feigenbaum attractor \cite%
{robledo1}, \cite{robledo3}.

\begin{figure}[h!]
\centering
\includegraphics[width=.5\textwidth]{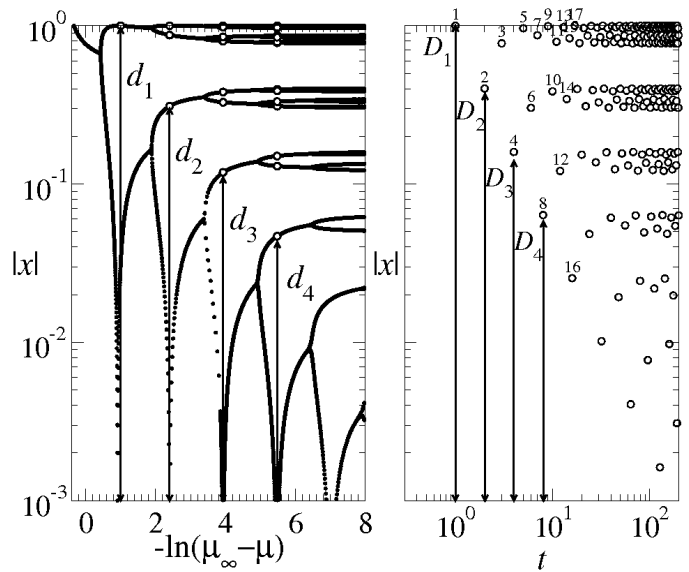}
\caption{
{\small \ Left panel: Absolute value of attractor positions for the
logistic map }${\small f}_{\mu }{\small (x)}${\small \ in logarithmic scale
as a function of the logarithm of the control parameter difference }${\small %
\mu }_{\infty }{\small -\mu }${\small . Right panel: Absolute value of
trajectory positions for the logistic map }${\small f}_{\mu }{\small (x)}$%
{\small \ at }${\small \mu }_{\infty }${\small \ with initial condition }$%
{\small x}_{0}{\small =0}${\small \ in logarithmic scale as a function of
the logarithm of time }$t${\small , also show by the numbers close to the
circles. The arrows indicate the equivalence between the diameters }$d_{N}$%
{\small \ in the left panel, and position differences }$D_{N}${\small \ with
respect to }${\small x}_{0}{\small =0}${\small \ in the right panel.}
}
\label{fig1}
\end{figure}

\section{Preimage structure of attractor and repellor}

\begin{figure}[h!]
\centering
\includegraphics[width=.5\textwidth]{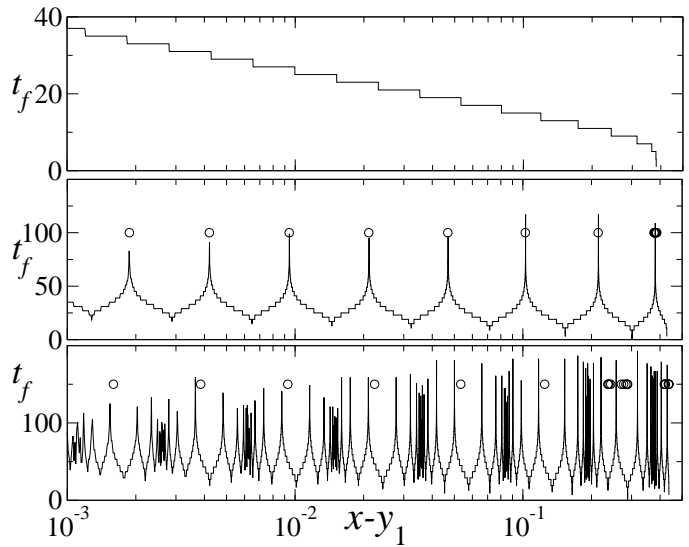}
\caption{
{\small Time of flight }${\small t}_{f}{\small (x)}${\small , the
number of iterations necessary for a trajectory with initial condition at }$%
{\small x}${\small \ to reach an attractor position as a function of the
distance, in logarithmic scale, between }${\small x}${\small \ and the
repellor position at }${\small y}_{{\small 1}}${\small . The top panel is
for the twice-iterated map }$f_{\overline{\mu }_{1}}^{(2)}(x)${\small , }$%
\overline{\mu }_{1}={\small 1}${\small , }${\small y}_{{\small 1}}\simeq 
{\small 0.6180340}${\small .\ The middle panel for }$f_{\overline{\mu }%
_{2}}^{(4)}(x)${\small ,} $\overline{\mu }_{2}\simeq {\small 1.31070264}$%
{\small , }${\small y}_{{\small 1}}\simeq {\small 0.571663}${\small . And
the bottom panel for }${\small f}_{\overline{\mu }_{3}}^{(8)}{\small (x)}$%
{\small ,} $\overline{\mu }_{3}{\small \simeq 1.38154748}${\small , }$%
{\small y}_{{\small 1}}{\small \simeq 0.56264475}${\small . The values of }$%
{\small x}${\small \ near the peaks (open circles) correspond to initial
conditions very close to the repellor preimages. See text.}
}
\label{fig2}
\end{figure}

A convenient way to visualize how the preimages for the Feigenbaum attractor
and repellor are distributed and organized is to consider the simpler
arrangements for the preimages of the supercycles' attractors and repellors.
These arrangements have been characterized in detail recently \cite{moyano1}
for supercycles of small periods $2^{N}$, $N=1,2,\ldots$, and it was observed
how they become more complicated very rapidly as $N$ grows \cite{moyano1}.
For $N=1$ the preimages of the attractor belong to only two basins, one for
each position of the attractor, separated by the positions of the
fixed-point repellor and its single preimage. For $N=2$ the attractor
preimages are distributed into four basins, one pair of them separated from
the other pair by a fractal boundary built around the positions of the first
or `old' $N=1$ repellor and its preimage. This fractal boundary consists of
the infinite number of preimages that the old repellor has acquired and this
in turn cluster exponentially around it and around its `old' preimage. For $%
N=3$ the attractor preimages are distributed into eight basins, four of the
basins are separated from the remaining four by a more complex fractal
boundary sector, built again about the positions of the old $N=1$ repellor
and its old preimage. This sector is made of an infinite number of replicas
of the fractal boundary for the case with $N=2$, the increasingly more
numerous preimages of the old repellor form now an infinite set of clusters,
grouping themselves exponentially around it and around its old preimage. The
basins for the positions of the attractor are also separated from each other
in pairs via other fractal boundary segments with the characteristics of the 
$N=2$ case, structures that appear centered now around the next generation
of repellors (two) and their first preimages (two). In the panels of Fig. \ref{fig2}
we show the relative times $t_{f}$ required to reach the attractor as a
function of the initial condition $x_{0}$ for the supercycles with $N=1$, $2$%
, and $3$, respectively. The logarithmic horizontal scale in these figures
relative to the position of the old repellor reveals the main characteristic
of the preimage structure for each case. These figures illustrate our
description above. See Ref. \cite{moyano1} for the definition of $t_{f}$ and
further details.

As the period $2^{N}$ increases the preimage structures for the attractor
and repellor become more and more involved, with the appearance of new
features made up of an infinite repetition of building blocks each of which
is equivalent to the more dense structures present in the previous $2^{N-1}$
case. In addition all other structures in the earlier $2^{N-2}$, \ldots, $2^{1}$
cases are still present. Thus a hierarchical organization of preimages is
built upon as $N$ increases, and the preimage layout for the Feigenbaum
attractor and repellor is obtained as the limiting form of the rank
structure of the fractal boundaries between attractor basins, that are
formed by the preimages of the repellor positions for the families of
periodic attractors symbolized by the supercycles. The fractal boundaries
consist of sub-basins of preimages for the attractor positions separated by
preimages of the repellor positions. The sizes of these sub-basins decrease
while their numbers increase as $N\rightarrow \infty $, with the result that
in this limit their sizes vanish and the preimages of both attractor and
repellor positions become two sets - with dimension equal to the dimension
of phase space - dense in each other. (The attractor and repellor are two
multifractal sets with dimension $d_{f}\simeq 0.538\ldots$ \cite{schuster1}).
For the $2^{N}$-th supercycle the positions for its repellor and its first
preimage appear located at the inflection points of $f_{\overline{\mu }%
_{N}}^{(2^{N})}(x)$, and it is in the close vicinity of them that the
fractal boundaries of basins of attraction for the positions of the
attractor form. To illustrate how the dense sets of preimage structures for
the Feigenbaum attractor and repellor develop we plot in Fig. 3 the absolute
value of $df_{\overline{\mu }_{N}}^{(2^{N})}/dx$
for $N=1,2,\ldots,4$ vs. $x$. The maxima in this curve correspond to the
inflection points of $f_{\overline{\mu }_{N}}^{(2^{N})}(x)$ at which the
repellor positions or their first preimages are located.

\begin{figure}[h!]
\centering
\includegraphics[width=.5\textwidth]{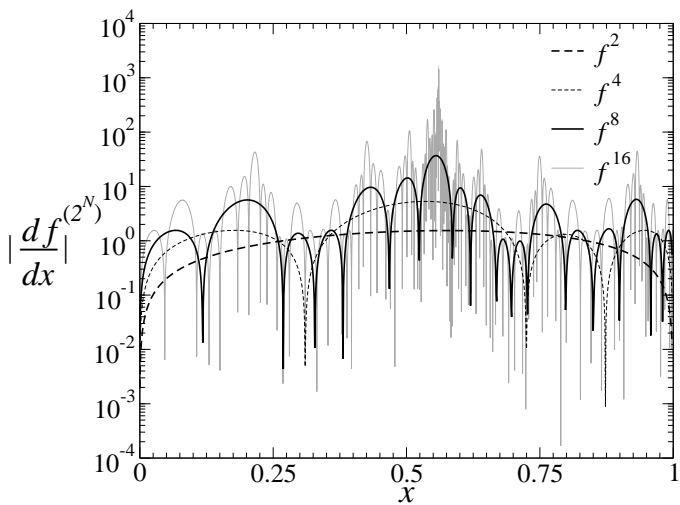}
\caption{
{\small The absolute value of }$df_{\overline{\mu }%
_{N}}^{(2^{N})}(x)/dx$,{\small \ for }${\small N=1,2,3}${\small \ and }$%
{\small 4}${\small , in logarithmic scale as a function of }${\small x}$%
{\small \ in the interval }${\small 0\leq x\leq 1}${\small . The
proliferation of maxima conveys the development of the hierarchical
structure of repellor preimages. See text. }
}
\label{fig3}
\end{figure}

\section{Sequential opening of phase space gaps}

One way wherein the preimage structure described above manifests in the
dynamics is via the successive formation of phase space gaps that ultimately
give rise to the attractor and repellor multifractal sets. In order to
observe explicitly this process we consider an ensamble of initial
conditions $x_{0}$ spread out uniformly across the interval $-1\leq
x_{0}\leq 1$ and keep track of their positions at subsequent times. In Figs.
4 to 6 we illustrate the outcome for the supercycles of periods $2^{2}$, $%
2^{3}$ and $2^{4}$, respectively, where we have plotted the time evolution
of an ensemble composed of $10000$ trajectories. In the
left panel of each figure we show the absolute value of the positions $%
\left\vert x_{t}\right\vert $ vs time $t$, while, for comparison purposes,
in the right panel we show the absolute value of $\left\vert x\right\vert $
both \textit{vs} $f_{\overline{\mu }_{N}}^{(2^{N})}(x)$ and \textit{vs} $%
\left\vert df_{\overline{\mu }_{N}}^{(2^{N})}/dx\right\vert $ to facilitate
identification of the attractor and repellor positions. The labels $%
k=1,2,3,\ldots$ indicate the order of the gap set (or equivalently the order of
the repellor generation set \cite{moyano1}). In Fig. 4 (with $\mu =\overline{%
\mu }_{2}$) one observes a large gap opening first that contains the old
repellor ($k=0$) in its middle region and two smaller gaps opening afterward
that contain the two repellors of second generation ($k=1$) once more around
the middle of them. In Fig. 5 (with $\mu =\overline{\mu }_{3}$) we initially
observe the opening of a primary and the two secondary gaps as in the
previous $\mu =\overline{\mu }_{2}$ case, but subsequently four new smaller
gaps open each around the third generation of repellor positions ($k=2$). In
Fig. 6 (with $\mu =\overline{\mu }_{4}$) we observe the same development as
before, however at longer times eight additional and yet smaller gaps emerge
each around the fourth generation of repellor positions ($k=3$). Naturally,
this process continues indefinitely as $N\rightarrow \infty $ and
illustrates the property mentioned before for $\mu _{\infty }$, that time
evolution at fixed control parameter value resembles progression from $\mu =0
$ up to, in this paragraph, $\overline{\mu }_{N}$. It is evident in all
Figs. 4 to 6 that the closer the initial conditions $x_{0}$ are to the
repellor positions the longer times it takes for the resultant trajectories
to clear the gap regions. This intuitively evident feature is essentially
linked to the knowledge we have gained about the fractal boundaries of the
preimage structure, and the observable `bent over' portions of these
distinct trajectories in the figures correspond to their passage across the
boundaries. (Since the ensemble used in the numerical experiments is finite
there appear only a few such trajectories in Figs. 4 to 6).

\begin{figure}[h!]
\centering
\includegraphics[width=.5\textwidth]{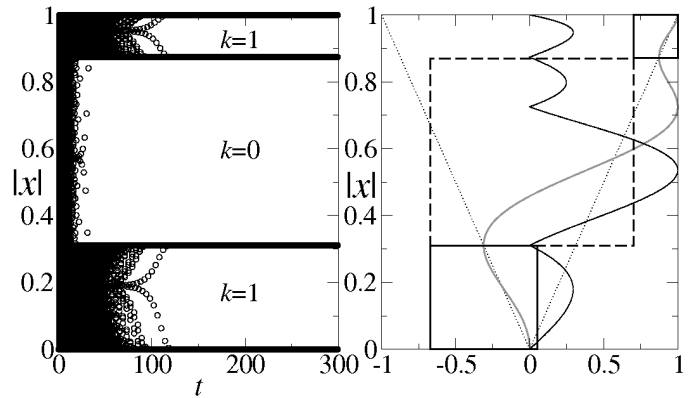}
\caption{
{\small \ Phase-space gap formation for }${\small \mu =}\overline{\mu 
}_{2}${\small . Left panel: time evolution of a uniform ensemble of 10000
trajectories as a function of }$|x|${\small \ (black areas and open
circles). The values of the index }${\small k}${\small \ label the order of
the gap set. Right panel: Turned around plots of }$f_{\overline{\mu }%
_{2}}^{(4)}(x)${\small (grey)\ and }$\left\vert df_{\overline{\mu }%
_{2}}^{(4)}(x)/dx\right\vert ${\small (black) \ vs }$|x|${\small \ as guides for the
identification of attractor and repellor positions.}
}
\label{fig4}
\end{figure}

To facilitate a visual comparison between the process of gap formation at $%
\mu _{\infty }$ and the dynamics inside the Feigenbaum attractor (as
illustrated by the trajectory in Fig. 1b) we plot in Fig. 7 the time
evolution of the same ensemble composed of $10000$
trajectories with $\mu =\mu _{\infty }$. We use this time logarithmic scales
for both $\left\vert x_{t}\right\vert $ and $t$ and then superpose on the
evolution of the ensemble the positions for the trajectory starting at $%
x_{0}=0$. It is clear from this figure that the larger gaps that form
consecutively  have all the same width in the logarithmic scale of the plot
and therefore their actual widths decrease as a power law, the same power
law followed, for instance, by the position subsequence $x_{t}=\alpha ^{-N}$%
, $t=2^{N}$, $N=0,1,2,\ldots$ for the trajectory inside the attractor starting
at $x_{0}=0$. This set of gaps develop in time beginning with the largest
one containing the $k=0$ repellor, then followed by a second gap, one of a
set of two gaps associated to the $k=1$ repellor, next a third gap, one gap
of a set of four gaps associated to the $k=2$ repellor, and so forth. The
locations of this specific family of consecutive gaps advance monotonically
towards the most sparse region of the multifractal attractor located at $x=0$%
. The remaining gaps formed at each stage converge, of course, to locations
near other regions of the multifractal but are not easily seen in Fig. 7
because of the specific way in which this has been plotted (and because of
the scale used). In Fig. 8 we plot the same data differently, with the
variable $\ln \left\vert x\right\vert $ replaced by $\ln \left\vert
1-x\right\vert $ where now another specific family of gaps, one for each
value of $k=0,1,2,\ldots$, appear all with the same width in the logarithmic
scale, their actual widths decrease now as $\alpha ^{-2N}$, $N=0,1,2,\ldots$
The locations of this second family of consecutive gaps advance
monotonically towards the most crowded region of the multifractal attractor
located at $x=1$. The time necessary for the formation of \ successive gaps
of order $k=0,1,2,\ldots$, increases as $2^{k}$ because the duration of
equivalent movements of the trajectories across the corresponding preimage
structures involve the $2^{k}$-th composed function $f_{\overline{\mu }%
_{N}}^{(2^{k})}(x)$.

\begin{figure}[h!]
\centering
\includegraphics[width=.5\textwidth]{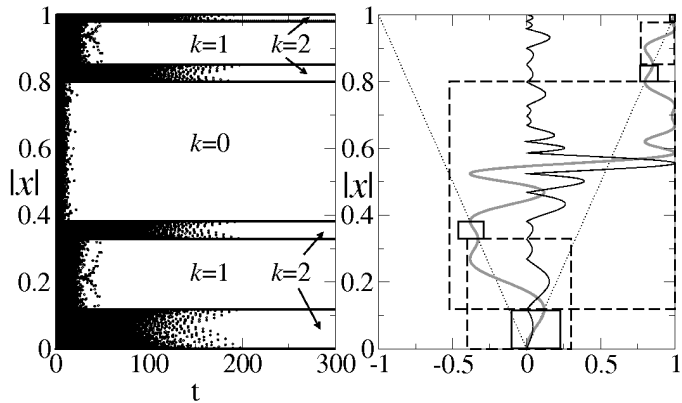}
\caption{
{\small Phase-space gap formation for }${\small \mu =}\overline{\mu }%
_{3}${\small . Left panel: time evolution of a uniform ensemble of 10000
trajectories as a function of }$|x|${\small \ (black areas and open
circles). The values of the index }${\small k}${\small \ label the order of
the gap set. Right panel: Turned around plots of }$f_{\overline{\mu }%
_{3}}^{(8)}(x)${\small (grey) \ and }$\left\vert df_{\overline{\mu }%
_{3}}^{(8)}(x)/dx\right\vert ${\small (black)\ vs }$|x|${\small \ as guides for the
identification of attractor and repellor positions.}
}
\label{fig5}
\end{figure}

\begin{figure}[h!]
\centering
\includegraphics[width=.5\textwidth]{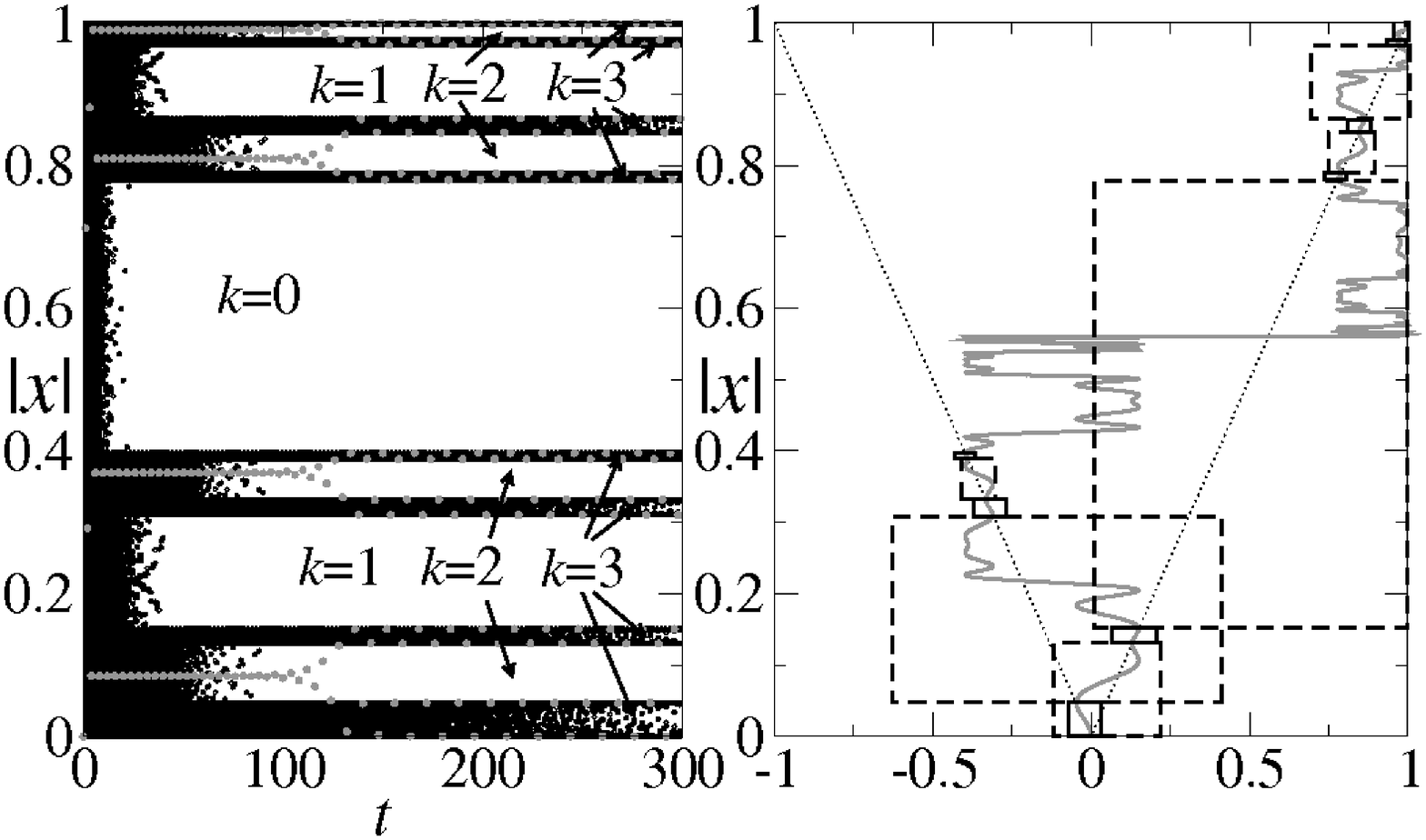}
\caption{
{\small Phase-space gap formation for }${\small \mu =\overline{\mu }%
_{4}}${\small . Left panel: time evolution of a uniform ensemble of 10000
trajectories as a function of }$|x|${\small \ (black areas and open
circles). The values of the index }${\small k}${\small \ label the order of
the gap set. Right panel: Turned around plots of }$f_{\overline{\mu }%
_{4}}^{(16)}(x)${\small (grey)\ and }$\left\vert df_{\overline{\mu }%
_{4}}^{(16)}(x)/dx\right\vert ${\small (black)\ vs }$|x|${\small \ as guides for
the identification of attractor and repellor positions.}
}
\label{fig6}
\end{figure}

\begin{figure}[h!]
\centering
\includegraphics[width=.5\textwidth]{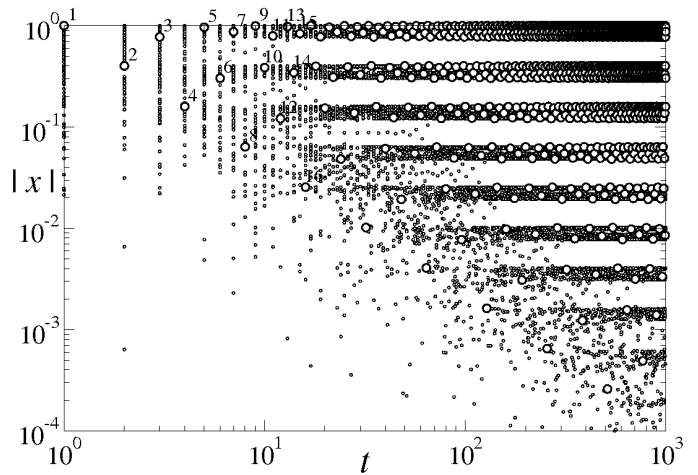}
\caption{
{\small Phase-space gap formation for }$\mu ={\small \mu }_{\infty }$%
{\small . The black dots correspond to time evolution of a uniform ensemble
of 10000 trajectories as a function of }$|x|${\small \ vs }${\small t}$%
{\small ,\ both in logarithmic scales. The open circles are the positions,
labeled by the times at which they are reached, for the trajectory inside
the Feigenbaum attractor with initial condition }${\small x}_{{\small 0}}%
{\small =0}${\small , same as right panel in Fig. 1.}
}
\label{fig7}
\end{figure}

\begin{figure}[h!]
\centering
\includegraphics[width=.5\textwidth]{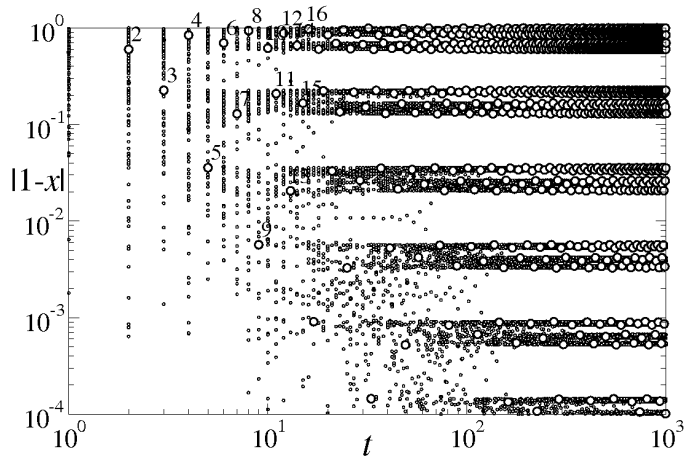}
\caption{
{\small Same as Fig. 7 but with replacement of }$|x|${\small \ by }$|%
{\small 1}-x|${\small . Notice the change in slope with respect to Fig. 7 in
the opening of gaps and in the layout of the positions for the trajectory
inside the attractor. }
}
\label{fig8}
\end{figure}

\section{Scaling for the rate of convergence to the attractor and repellor}

\begin{figure}[h!]
\centering
\includegraphics[width=.5\textwidth]{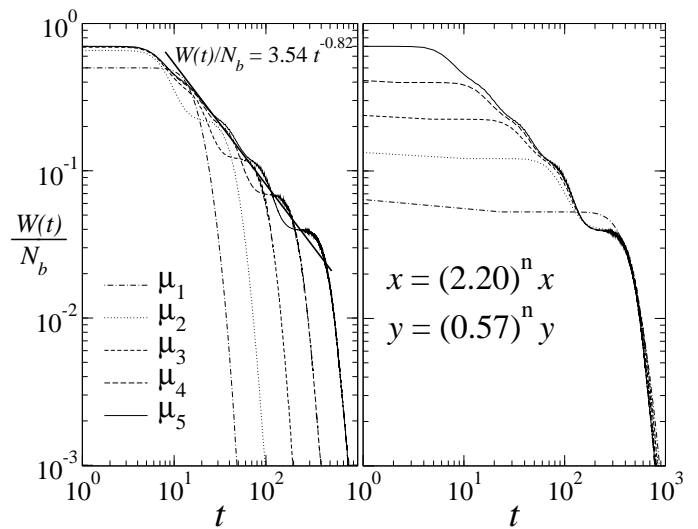}
\caption{
{\small Left panel: The rate }${\small W(t)}${\small , divided by
the number of boxes }${\small N}_{b}${\small \ employed,\ of approach to the
attractor for the supercycles of periods }${\small 2}^{{\small N}}${\small , 
}${\small N=1,2,3,4}${\small \ and }${\small 5}${\small \ in logarithmic
scales. The expression shown corresponds to the power-law decay of the
developing logarithmic oscillations. Right panel: Superposition of the five
curves for }${\small W(t)}${\small \ in the left panel via }${\small n}$%
{\small -times repeated rescaling factors shown for the horizontal }${\small %
x}${\small \ and vertical }${\small y}${\small \ axis.}
}
\label{fig9}
\end{figure}

There is \cite{lyra1} an all-inclusive and uncomplicated way to measure the
rate of convergence of an ensemble of trajectories to the attractor (and to
the repellor) that consists of a single time-dependent quantity. A partition
of phase space is made of $N_{b}$ equally sized boxes or bins and a uniform
distribution, of $N_{c}$\ initial conditions placed along the interval $%
-1\leq x\leq 1$, is considered again. The number $r$ of trajectories per box
is $r=N_{c}/N_{b}$. The quantity of interest is the number of boxes $W(t)$
that contain trajectories at time $t$. This is shown in Fig. 9 in
logarithmic scales for the first five supercycles of periods $2^{1}$ to $%
2^{5}$ where we can observe the following features: In all cases $W(t)$
shows a similar initial nearly constant plateau and a final well-defined
decay to cero. As it can be observed in the left panel of Fig. \ref{fig9} 
the duration of the final
decay grows (approximately) proportional to the period $2^{N}$ of the
supercycle. There is an intermediate slow decay of $W(t)$ that develops as $N
$ increases with duration also (just about) proportional to $2^{N}$. For the
shortest period $2^{1}$ there is no intermediate feature in $W(t)$, this
appears first for period $2^{2}$ as a single dip and expands with one
undulation every time $N$ increases by one unit. The expanding intermediate
regime exhibits the development of a power-law decay with the logarithmic
oscillations characteristic of discrete scale invariance \cite{sornette1}.
Clearly, the manifestation of discrete invariance is expected to be
associated to the period-doubling cascade. In the right panel of Fig. \ref{fig9} 
we show a superposition of the five curves in Fig. \ref{fig9} (left panel) 
obtained via rescaling of both $W(t)$ and $t$ for each curve according to repeated scale factors.

\begin{figure}[h!]
\centering
\includegraphics[width=.5\textwidth]{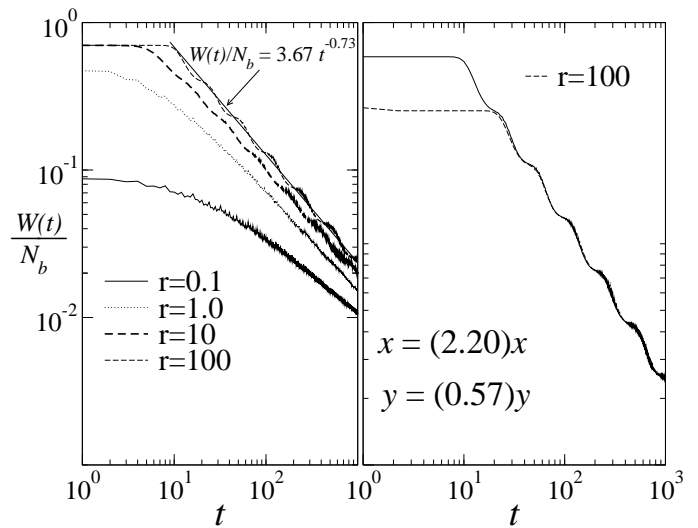}
\caption{
{\small Left panel: The rate }${\small W(t)}${\small \ of approach
to the Feigenbaum attractor at }$\mu ={\small \mu }_{\infty }${\small \ in
logarithmic scales. The curves correspond to the values given for the number 
}${\small r}${\small \ of trayectories per box, and the expression shown
corresponds to the power-law decay with logarithmic oscillations. Right
panel: Superposition of }${\small W(t)}${\small , for }${\small r=100}$%
{\small , with itself via the rescaling shown (the same as in Fig. 8) for
the horizontal }${\small x}${\small \ and vertical }${\small y}${\small \
axis.}
}
\label{fig10}
\end{figure}

\begin{figure}[h!]
\centering
\includegraphics[width=.5\textwidth]{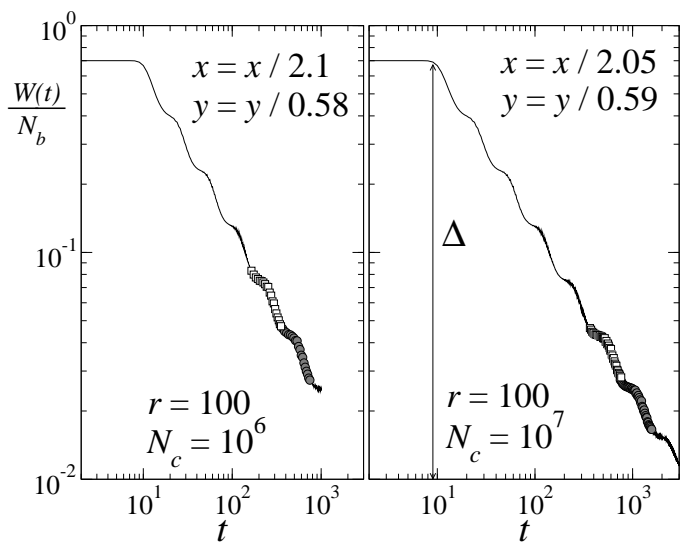}
\caption{
{\small The same as in the right panel of Fig. 10 but obtained with
increased precision. In the left panel the number of initial conditions is }$%
{\small N}_{c}{\small =10}^{6}${\small , while in the right panel }${\small N%
}_{c}{\small =10}^{7}{\small }${\small .} {\small The distance} $\Delta$ {\small is }
$\Delta= (1+|-1/\alpha|)/(1+|-1|)$, {\small where }$\alpha${ is Feigenbaum's constant. 
This stems from the fact that all initial conditions out of the interval} $(-1/ \alpha, 1)$ 
{\small take a value inside this interval in the first iteration.} {\small As it can be observed the
scaling factor for the horizontal axis converges to the exact value }$%
{\small x=2}${\small .}
}
\label{fig11}
\end{figure}

The limiting form $W(t)$ for $N\rightarrow \infty $ is shown in the left 
panel of Fig. \ref{fig10} for various values of $r$ while in its right panel we show, 
for $r=100$, a scale amplification of $W(t)$ with the same
factors employed in Fig. \ref{fig9} for the supercycles with small periods. The 
behavior of $W(t)$ at $\mu _{\infty }$ was originally presented in Ref. \cite%
{lyra1} where the power law exponent $s$ and the logarithmic oscillation
parameter $\Lambda $ in%
\begin{equation}
W(t)\simeq h\left( \frac{\ln t}{\ln \Lambda }\right) t^{-s}
\end{equation}%
were obtained numerically with a precision that corresponds to $r=10$. In
Eq. (2) $h(x)$ is a periodic function and $\Lambda $ is the scaling factor
between the periods of two consecutive oscillations. More recently, in Ref. 
\cite{grassberger1} it was pointed out that numerical estimates of $W(t)$
are subject to large finite-size corrections, and, also, that $W(t)$ should
scale with the intervals in the triadic cantor set construction of the
Feigenbaum attractor \cite{grassberger1}\cite{comment},  from which the value for $s\cong
0.800138194$ is reported. The values for the rescaling factors in our Figs. 
\ref{fig9} and \ref{fig10} suffer from these large finite size 
effects due to the relatively small values of $r$ used 
in the calculations. This is evident since the time
scaling factor obtained from these data differs $10\%$ from the exact value
of $\Lambda =2$ implied by the discrete scale invariance property created by
the period-doubling cascade. In Fig. \ref{fig11} we show the rate $W(t)$ and the
superposition of repeated amplifications of itself (as in the right panel of 
Fig. \ref{fig10}) for increasing values of $N_c$. 
We find that the scaling factor $\Lambda $ converges to its limit $\Lambda =2$.

We are now in a position to appreciate the dynamical mechanism at work
behind the features of the decay\ rate $W(t)$. From our previous discussion
we know that every time the period of a supercycle increases from $2^{N-1}$
to $2^{N}$ by shifting the control parameter value from $\overline{\mu }%
_{N-1}$ to $\overline{\mu }_{N}$, the preimage structure advances one stage
of complication in their hierarchy. Along with this, and in relation to the
time evolution of the ensemble of trajectories, an additional set of $2^{N}$
smaller phase-space gaps develops and also a further oscillation takes place
in the corresponding rate $W(t)$ for finite-period attractors. At $\mu =\mu
_{\infty }$ time evolution tracks the period-doubling cascade progression
and the flow of trajectories undergo every time $t$ increases from $2^{N-1}$
to $2^{N}$ equivalent passages across stages in their itinerary through the
preimage ladder structure, in the development of phase-space gaps, and in
logarithmic oscillations in $W(t)$. In Fig. 12 we show the correspondence
between the latter features quantitatively. Also, in this figure we have
added the results of a calculation of $W(t)$ at times $t=2^{N}$, $N=0,1,2,\ldots
$, according to the expression%
\begin{equation}
W(2^{N})=\sum_{m=1}^{N}d_{N,m}.
\end{equation}%
This confirms the value $s\cong 0.8001$.

\begin{figure}[h!]
\centering
\includegraphics[width=.5\textwidth]{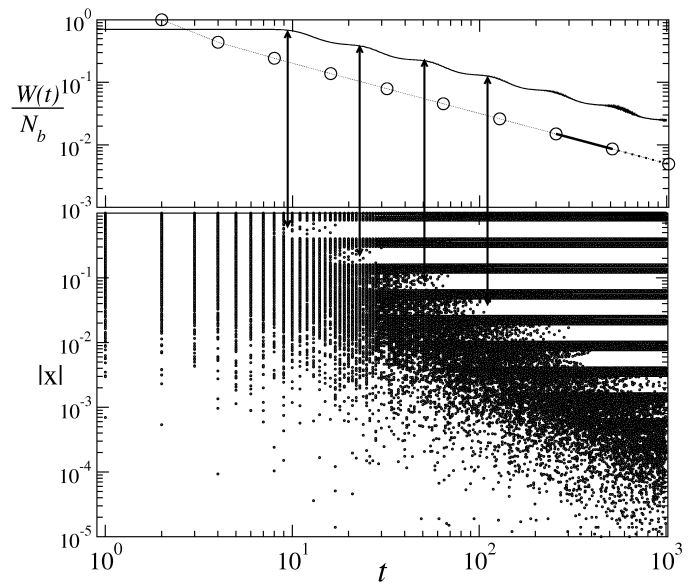}
\caption{
{\small Correspondence between the power-law decay with
log-periodic oscillation features of the rate }${\small W(t)}${\small \ with
the sequential opening of phase space gaps. Top panel: The solid line is }$%
{\small W(t)}${\small \ from Fig. 9 and the open circles values are obtained
for }${\small W(t)}$ {\small from Eq. (3) at times }$t={\small 2}^{{\small N}%
}${\small , }$N=1,2,\ldots${\small \ See text.}
}
\label{fig12}
\end{figure}

\section{Summary}

We have examined the process followed by an ensemble of uniformly
distributed initial conditions $x_{0}$ across the phase space to arrive at
the Feigenbaum attractor, or get captured by its corresponding repellor.
Significantly, we have gained understanding concerning the dynamical
ordering in the $x_{0}$, in relation to the construction of the families of
phase-space gaps that support the attractor and repellor, and about the rate
of approach of trajectories towards these multifractal sets, as measured by
the fraction of bins $W(t)$ still occupied by trajectories at time $t$. An
important factor in obtaining this knowledge has been the consideration of
the equivalent dynamical properties for the supercycles of small periods in
the bifurcation cascade \cite{moyano1}.

As we have seen, a doubling of the period introduces well-defined additional
elements in the hierarchy of the preimage structure, in the family of phase
space gaps, and in the log-periodic power law decay of the rate $W(t)$. We
have then corroborated the wide-ranging correlation between time evolution
at $\mu _{\infty }$ from $t=0$ up to $t\rightarrow \infty $ with the
`static' period-doubling cascade progression from $\mu =0$ up to $\mu
_{\infty }$. As a result of this we have acquired an objective insight into
the complex dynamical phenomena that fix the decay\ rate $W(t)$.\ We have
clarified the genuine mechanism by means of which the discrete scale
invariance implied by the log-periodic property in $W(t)$ arises, that is,
we have seen how its self-similarity originates in the infinite hierarchy
formed by the preimage structure of the attractor and repellor. The rate $%
W(t)$ can be obtained quantitatively (see Eq. (3)) from the supercycle
diameters $d_{N,m}$. This basic data descriptive of the period-doubling
route to chaos is also a sufficient ingredient in the determination of the
anomalous sensitivity to initial conditions for the dynamics inside the
Feigenbaum attractor \cite{robledo3}.

\textbf{Acknowledgments.} Partial support by DGAPA-UNAM and CONACyT (Mexican
Agencies) is acknowledged.


\begin{thebibliography}{99}
\bibitem{comment} Evidently, the cantor set construction of the Feigenbaum attractor 
is comparable to our description of consecutive gap formation.
\end{thebibliography}
\end{document}